\documentclass[smallextended]{svjour3}       
\smartqed  
\usepackage[numbers]{natbib}
\usepackage{graphicx}
\usepackage{dcolumn}
\usepackage{bm}
\usepackage{amssymb}
\usepackage{subfigure}
\usepackage{mathtools}

\makeatletter   
\newcommand*{\rom}[1]{\expandafter\@slowromancap\romannumeral #1@}
\makeatother

\newtheorem{defn}{Definition}

\usepackage[normalem]{ulem}

\usepackage[table]{xcolor} 
\usepackage{array}
\usepackage{tabularx}
\usepackage{tikz}
\usepackage{lipsum}
\usetikzlibrary{calc,shadings,patterns}

\makeatletter
\tikzset{%
  remember picture with id/.style={%
    remember picture,
    overlay,
    save picture id=#1,
  },
  save picture id/.code={%
    \edef\pgf@temp{#1}%
    \immediate\write\pgfutil@auxout{%
      \noexpand\savepointas{\pgf@temp}{\pgfpictureid}}%
  },
  if picture id/.code args={#1#2#3}{%
    \@ifundefined{save@pt@#1}{%
      \pgfkeysalso{#3}%
    }{
      \pgfkeysalso{#2}%
    }
  }
}

\def\savepointas#1#2{%
  \expandafter\gdef\csname save@pt@#1\endcsname{#2}%
}

\def\tmk@labeldef#1,#2\@nil{%
  \def\tmk@label{#1}%
  \def\tmk@def{#2}%
}

\tikzdeclarecoordinatesystem{pic}{%
  \pgfutil@in@,{#1}%
  \ifpgfutil@in@%
    \tmk@labeldef#1\@nil
  \else
    \tmk@labeldef#1,(0pt,0pt)\@nil
  \fi
  \@ifundefined{save@pt@\tmk@label}{%
    \tikz@scan@one@point\pgfutil@firstofone\tmk@def
  }{%
  \pgfsys@getposition{\csname save@pt@\tmk@label\endcsname}\save@orig@pic%
  \pgfsys@getposition{\pgfpictureid}\save@this@pic%
  \pgf@process{\pgfpointorigin\save@this@pic}%
  \pgf@xa=\pgf@x
  \pgf@ya=\pgf@y
  \pgf@process{\pgfpointorigin\save@orig@pic}%
  \advance\pgf@x by -\pgf@xa
  \advance\pgf@y by -\pgf@ya
  }%
}
\newcommand\tikzmark[2][]{%
\tikz[remember picture with id=#2] {#1;}}
\makeatother

\newcommand\HatchedCell[4][0pt]{%
  \begin{tikzpicture}[overlay,remember picture]%
    \fill[#4] ( $ (pic cs:#2) + (0,1.9ex) $ ) rectangle ( $ (pic cs:#3) + (0pt,-#1*\baselineskip-.8ex) $ );
  \end{tikzpicture}%
}%

\begin{document}

\title{Is contextuality about the identity of random
variables?
}


\author{Mojtaba Aliakbarzadeh        \and
        Kirsty Kitto 
}


\institute{M. Aliakbarzadeh \at
               School of Chemistry and Physics \\
       Queensland University of Technology\\
       Brisbane, 4000, Australia \\
              \email{m.aliakbarzadeh@qut.edu.au}           
           \and
           K. Kitto \at
              Connected Intelligence Centre\\
		University of Technology Sydney\\
        PO Box 123, Broadway, 2007, Australia
}

\date{Received: date / Accepted: date}

\maketitle

\begin{abstract}

Recent years have seen new general notions of contextuality emerge. Most of these employ \emph{context-independent} symbols to represent random variables in different contexts. As an example, the operational theory of \citet{spekkens2005contextuality} treats an observable being measured in two different contexts identically. Non-contextuality in this approach is the impossibility of drawing ontological distinctions between identical elements of the operational theory. However, a recent collection of work seeks to exploit \emph{context-dependent} symbols of random variables to interpret contextuality
\cite{KujalaDzhafarov2015Necessary,DzhafarovKujala2014ContextualityIdentity}. This approach associates contextuality with the possibility of imposing a particular joint distribution on random variables recorded under different experimental contexts. This paper compares these two different treatments of random variables and highlights the limitations of the context-dependent approach as a physical theory.
\end{abstract}

\maketitle

\section{Introduction}\label{Sec:Introduction}

Contextuality in quantum mechanics (QM) refers to the dependence of measurement results for specific observables upon the experimental arrangement being used to measure that observable
\citep{KochenSimon}. Although contextuality has been part of the conceptual framework of QM for decades, recent literature has attempted to arrive at a deeper understanding of this subtle concept. For example, \citet{Abramsky2011} unify the concepts of nonlocality and contextuality using sheaf theory,  \citet{CabelloSeveriniWinter2014Graph} use a graph theoretical approach to model contextuality, and  \citet{AcinFritz2015ACombinatorial}
construct a general contextuality model using the combinatorics of hypergraphs which generalises both the sheaf and graph theoretical approaches. Importantly to the argument mounted here, in 2005, \citet{spekkens2005contextuality} generalized the standard treatment of contextuality in QM to arbitrary operational theories, which allows for the identification of contextuality in theory-independent frameworks. 
These approaches have tried to unify our understanding of contextuality, however, each uses different mathematical structures and notations, making  comparison  difficult.  This  makes  it essential that we start to construct connections between them to improve our understanding of contextuality. Here we formally compare Spekkens' generalized notion of contextuality with a recent competing generalized notion of contextuality called Contextuality-by-Default (CbD) \cite{KujalaDzhafarov2015Necessary,DzhafarovKujala2014ContextualityIdentity}, which exploits \emph{context-dependent} symbols of random variables to interpret contextuality. Based on the comparison of these two approaches,  we specify the limitations of the CbD approach.

In Spekkens' approach, contextuality is defined as the non-existence of a statistically equivalent description at the ontological level for operationally equivalent procedures. However, in an actual experiment, it is not possible to attain exact operational equivalence \citep{MazurekPusey2016ExperimentaContextuality}. To solve this problem,  \citet{MazurekPusey2016ExperimentaContextuality} suggest a general method which considers equivalences not between the procedures, but certain convex combinations of them. Interestingly, an \emph{inexact operational equivalence} can also be achieved using the CbD notation. We show here that this is a result of the \emph{context-dependent} symbols of random variables. However, we indicate that this different realization of random variables does not provide a clear definition of ontic states. We point out that CbD is 
incapable of representing constraints on the joint measurability that one could realize between quantum observables in Specker's scenario. We also show that for a system satisfying non-signaling and no-disturbance conditions, CbD has to convert to the normal representation of random variables to satisfy the expected behaviour of that system. 

In Section~\ref{Sec:Preliminaries}, we briefly introduce the operational approach of Spekkens and the CbD approach. This is followed by comparison of these two approaches in Section \ref{Sec:ComparingTheApproaches}. Section~\ref{Sec:ProbabilitySpace} demonstrates how a different definition of probability space and random variables in the CbD approach lead to an inexact operational equivalence. Section~\ref{Sec:ParameterNon-signaling} investigates the differences between these two approaches for concepts like parameter independence and non-signaling conditions. And finally,  Section~\ref{Sec:CyclicExamples} evaluates the CbD approach using cyclic examples of contextuality.

\section{Preliminaries}\label{Sec:Preliminaries}

\subsection{Spekkens' approach }\label{Sec:Spekkens}

Spekkens' approach \cite{spekkens2005contextuality} uses an indeterministic ontological model that is more general than deterministic hidden variable models of contextuality. In the deterministic hidden variable models, the outcomes of measurements are determined by a given ontic state of the system (e.g. \citet{Fine1982Hidden}).
But in Spekkens' approach, the probabilities of the different outcomes of the measurement are determined by the ontic state. This is represented by the indicator function $\xi_{\lambda}(k|M)$ which is the probability distribution of the incidence of a measurement outcome $k$ given by implementing a measurement procedure $M$ for any ontic state $\lambda$.

In Spekkens' approach \cite{spekkens2005contextuality}, $\mu_{P}(\lambda)$ is the probability distribution of selecting the ontic states $\lambda \in \Lambda$ by a preparation procedure $P$. Here, two preparation procedures are operationally equivalent ($P \backsimeq P'$) if:
\begin{equation}{\label{equvalencepreparation}}
\begin{split}
p(k|M, P)=p(k|M, P') \quad \mbox{for all M}.
\end{split}
\end{equation}
Similarly, two measurement events are operationally equivalent ($[k|M]\backsimeq [K'|M']$) if:
\begin{equation}{\label{equivalentmeasurement}}
\begin{split}
p(k|M, P)=p(k'  |M', P) \quad \mbox{for all P}.
\end{split}
\end{equation}

Spekkens defines noncontextualty based on the definition of operational equivalence as:
\begin{defn}
 An ontological model is preparation noncontextual if we can represent every preparation procedure independent of context:
\begin{equation}{\label{PreparationNoncontextuality}}
\begin{split}
P \backsimeq P' \Rightarrow \mu_{P}(\lambda)=\mu_{P'}(\lambda) \quad \forall \lambda \in \Lambda,
\end{split}
\end{equation}
And the model is measurement noncontextual if we can represent every measurement event independent of context:
\begin{equation}{\label{MeasuremntNoncontextuality}}
\begin{split}
[k|M]\backsimeq [K'|M'] \Rightarrow \xi_{\lambda}(k|M)=\xi_{\lambda}(k|M') \quad \forall \lambda \in \Lambda.
\end{split}
\end{equation}
\end{defn}

\subsection{Contextuality-by-default}\label{Sec:CBD}

Contextuality-by-Default (CbD)
\cite{KujalaDzhafarov2015Necessary,DzhafarovKujala2014ContextualityIdentity,DzhafarovKujalaContext-Content2016} exploits  \emph{context-dependent} symbols of random variables to formalize contextuality. In this approach, random variables are represented using double indexing (e.g. $a_{q}^{c}$), where $q$ represents an observable (a physical property that we measure) and $c$ indicates a context of that measurement. In this model, a system of random variables comprises \emph{stochastically unrelated} ``bunches'', each of which is a set of jointly distributed random variables with the same context. The term ``stochastically unrelated'' is used to indicate that there is no joint distribution for the random variables when each random variable belongs to a different bunch.  

The CbD approach defines contextuality as the impossibility of imposing a joint distribution on  stochastically unrelated bunches. This imposed distribution is named  a  ``maximally connected coupling'' \citep{DzhafarovKujalaLarsson2015} which will be explained in more detail in Section \ref{Sec:CyclicExamples}.

\section{Comparing the approaches}\label{Sec:ComparingTheApproaches}

CbD initially emerged within the field of psychology \cite{DzhafarovKujala2012Selectivity}, but claims have since been made about its generality to the field of physics \cite{KujalaDzhafarov2015Necessary,DzhafarovKujala2014ContextualityIdentity}. This claim deserves some close scrutiny --- how does CbD compare with the results about contextuality that have emerged in the foundations of physics? 
We can start to see how the assumptions of the CbD approach subtly differ from those in the physics community with a consideration of previous work. For example, \citet{Shimony1984} defines a probability distribution $p_{\lambda}(a|\mathbb{A})$ (similar to $\xi_{\lambda}(a|\mathbb{A})$ in Spekkens' notation) on the set of ontic states in Bell's experiment, pointing to the impossibility of constructing a joint probability for non-commuting observables $\mathbb{A}_{1}$ and $\mathbb{A}_{2}$. This is similar to what CbD defines as \textit{stochastically unrelated} for two random variables $a_{\mathbb{A}_{1}}^{(\mathbb{A}_{1}, \mathbb{B}_{1})}$ and $a_{\mathbb{A}_{2}}^{(\mathbb{A}_{2}, \mathbb{B}_{1})}$. However, in the CbD method, two random variables $a_{\mathbb{A}_{1}}^{(\mathbb{A}_{1}, \mathbb{B}_{1})}$ and $a_{\mathbb{A}_{1}}^{(\mathbb{A}_{1}, \mathbb{B}_{2})}$ are defined as \textit{stochastically unrelated} as well, a situation for which there is no counterpart in Shimony's approach. This suggests that the theory might diverge from standard physical models and that a detailed comparison is warranted. In what follows we will compare the approach of CbD with that of Spekkens. 

\subsection{Merely close to operationally equivalent
}\label{Sec:ProbabilitySpace} 

Contextuality can emerge from non-commutativity of quantum observables, where the corresponding random variables of the non-commuting observables cannot be treated in a classical probability theory, since they cannot have a value at the same time. However, the CbD approach can be considered as a model within the framework of Kolmogorovian probability theory \citep{DzhafarovKujalaContext-Content2016,DzhafarovKon2018OnUniversalityOfClassical}. As \citet{DeBarrosKujala2015} state, to define the double-indexed random variables, we need a separate probability space for each possible context. Thus, a double-indexed random variable is defined as  $a_{i}^{j}: \Omega_{j} \rightarrow E_{i}$ \footnote{For simplicity, from now on we will use  $a_{i}^{j}$ instead of $a_{q_{i}}^{c_{j}}$, where subscripts $i$ indicate different observables and $j$ indicate different contexts.}, where  $E_{i}$ is the set of possible values related to an observable $i$, and $\Omega_{i}$ is the probability space related to a context $j$. As an example, for the observable $\mathbb{A}_{1}$ in Bell's experiment, $\Omega$ can be related to one of the two possible contexts $\{\mathbb{A}_{1},\mathbb{B}_{1}\}$ and $\{\mathbb{A}_{1},\mathbb{B}_{2}\}$. 

This consideration of different probability spaces or different random variables for only one observable in different contexts is not allowed within the definition of measurement contextuality suggested by Spekkens. In his model, the measurement procedures which admit contextuality on the ontological level are operationally context-independent. This was explained by \citet{SimmonsWallman2017ContextualityUnder}: 
\begin{quote}
	\emph{
	... the same notation is used for the objects in the first place, as a context-independent symbol is all that is needed to calculate probabilities.  However there is no formal argument to be made that these elements which are operationally context-independent should also be ontologically context-independent
	... (p.2)}
\end{quote}

 The double indexing notation associates e.g. two random variables $a_{\mathbb{A}_{1}}^{(\mathbb{A}_{1}, \mathbb{B}_{1})}$ and $a_{\mathbb{A}_{1}}^{(\mathbb{A}_{1}, \mathbb{B}_{2})}$ with the observable $\mathbb{A}_{1}$, where each different random variable is defined based on a different probability space. Substituting these two random variables instead of the two outcomes in the operational equivalence equation \eqref{equivalentmeasurement}, we obtain: $p(a_{\mathbb{A}_{1}}^{(\mathbb{A}_{1}, \mathbb{B}_{1})}  |\{(\mathbb{A}_{1}, \mathbb{B}_{1})\}, P)=p(a_{\mathbb{A}_{1}}^{(\mathbb{A}_{1}, \mathbb{B}_{2})}  |\{(\mathbb{A}_{1}, \mathbb{B}_{2})\}, P)$. This does not completely match with the original definition of operational equivalence. \citet{MazurekPusey2016ExperimentaContextuality} describe this new equation as: ``merely close to operationally equivalent''.

\subsection{Signaling conditions}\label{Sec:ParameterNon-signaling}

CbD suggests a measure of contextuality for both the case of signaling and non-signaling \cite{DzhafarovKujalaLarsson2015}. In this section we focus on constraints for signaling (the Parameter independence (PI) and non-signaling conditions \cite{Jarrett1984, Shimony1984Controllable, Maudlin2002}),  investigating their possible representation using the CbD notation.

First, consider two \textit{stochastically unrelated} random variables $a_{\mathbb{A}_{1}}^{(\mathbb{A}_{1}, \mathbb{B}_{1})}$ and $a_{\mathbb{A}_{1}}^{(\mathbb{A}_{1}, \mathbb{B}_{2})}$. At a superficial level, we may assume that the PI condition for each ontic state $\lambda$ is satisfied if $\xi_{\lambda}(a_{\mathbb{A}_{1}}^{(\mathbb{A}_{1}, \mathbb{B}_{1})}|\mathbb{A}_{1}, \mathbb{B}_{1}) = \xi_{\lambda}(a_{\mathbb{A}_{1}}^{(\mathbb{A}_{1}, \mathbb{B}_{2})}|\mathbb{A}_{1}, \mathbb{B}_{2})$. But we cannot check the validity of this representation since the CbD approach does not have a clear position about the ontic state. 
\citet{KujalaDzhafarov2015Necessary} discuss the existence of joint distribution and its relation to a hidden variable $\lambda$:
\begin{quote}
\emph{The existence of a joint distribution of several random variables is equivalent to the possibility of presenting them as functions of a single, hidden variable $\lambda$}.
\end{quote}

But this fails to provide a more specific definition of precisely how the \emph{contexts} of the double indexed random variables relate to $\lambda$. We can discuss this further by comparing CbD with other contextual models. In Sppekens' approach, the ontic state of the system is specified by hidden variables. If we assume that the hidden variables in this quote are the same as the ontic states in Spekkens' approach, still we cannot consider CbD as an indeterministic ontological model (see Section~\ref{Sec:Spekkens}), because CbD does not associate probabilities such as $\xi_{\lambda}(k|M)$ to a given ontic state of the system. Instead, in CbD the outcomes of the measurements are determined as \emph{functions} of $\lambda$. Therefore, comparing CbD with deterministic \emph{contextual} hidden variable models could help us better to find out how the \emph{contexts} of the double indexed random variables relate to $\lambda$. As an example, we employ Fine's deterministic contextual hidden variable model for Bell's experiment \cite{Fine1982Hidden,Shimony1984}. This represents the complete state specification $\Lambda$ as a set of sixteen quadruples $\lambda = \langle a_{1}, a_{2}, b_{1}, b_{2} \rangle$, where $a_{i}, b_{i}=\pm 1$, and with response functions $A_{1}(\lambda)= a_{1}$, $A_{2}(\lambda)= a_{2}$, $B_{1}(\lambda)= b_{1}$ and $B_{2}(\lambda)= b_{2}$. Now we can assume that the '\emph{functions}' in the above quote are the same as the response functions as occur in Fine's model. We can assume further that the CbD considers a single hidden variable $\lambda$ for the two possible contexts of each random variable such as $a_{1}$. This would lead to a $\Lambda$ with 256 octuples $\lambda = \langle a_{1}^1, a_{1}^2, a_{2}^1, a_{2}^2, b_{1}^1, b_{1}^2, b_{2}^1, b_{2}^2 \rangle$. But if one wanted to use the CbD notation, it is not clear how precisely they might convince themselves to not use two distinct hidden variables $\lambda = \langle a_{1}^1, b_{1}^1 \rangle$ and  $\lambda' = \langle a_{1}^2, b_{2}^1 \rangle$ for the two different contexts of the random variable $a_{1}$. In other words, when the \emph{notation} of random variables itself carries the meaning of contextuality, why should we still use contextual hidden variables? We should note that this choice of hidden variables could lead to even more possible elements for $\Lambda$. However, we believe either of these two possible choices of the deterministic hidden variables based on the CbD notation only adds unnecessary complexity without adding anything to our understanding of contextuality as it arises in physics.

Unlike the PI condition, the non-signaling condition is expressed independently of the ontic state $\lambda$. This provides an opportunity to precisely explore the meaning of this condition in CbD and compare it with other approaches.
In Table \ref{CBDprobabilities}, we represent the joint probability distributions of Bell's experiment using the CbD notation. This table shows that double indexing can preserve the original meaning of non-signaling only when it converts to the standard \emph{context-independent} representation of random variables. As an example, the probability of $a_{\mathbb{A}_{1}}=+1$ is independent of the setting for measurement in the other side of the experiment if $p_{1}+ p_{2} = p_{3}+ p_{4}$. As a result of this equality, two random variables $a_{\mathbb{A}_{1}}^{(\mathbb{A}_{1}, \mathbb{B}_{1})}$ and $a_{\mathbb{A}_{1}}^{(\mathbb{A}_{1}, \mathbb{B}_{2})}$ have the same distribution for the same value of +1 \footnote{It is clear that one can check whether the two random variables $a_{\mathbb{A}_{1}}^{(\mathbb{A}_{1}, \mathbb{B}_{1})}$ and $a_{\mathbb{A}_{1}}^{(\mathbb{A}_{1}, \mathbb{B}_{2})}$ have the same distribution only when they have a same value.}, or in other words, they must have the same representation ($a_{i}$). 
\begin{table}[!hbtp]
   	\caption{The joint probability distributions for Bell's experiment using the double indexing scenario.}
   	    \label{CBDprobabilities}
   	    \centering
   	    \vspace*{.20 cm}
   	    {\setlength{\extrarowheight}{4pt}
\begin{tabular}{ccc}
		$\mathbb{A}_{1}, \mathbb{B}_{1}$ & $a_{\mathbb{B}_{1}}^{(\mathbb{A}_{1}, \mathbb{B}_{1})}= +1$ & $a_{\mathbb{B}_{1}}^{(\mathbb{A}_{1}, \mathbb{B}_{1})}=-1$ \\ [2ex]\cline{2-3} 
\multicolumn{1}{c|}{$a_{\mathbb{A}_{1}}^{(\mathbb{A}_{1}, \mathbb{B}_{1})}= +1$} & \multicolumn{1}{c|}{$p_{1}$} & \multicolumn{1}{c|}{$p_{2}$} \\ [2ex]\cline{2-3} 
\multicolumn{1}{c|}{$a_{\mathbb{A}_{1}}^{(\mathbb{A}_{1}, \mathbb{B}_{1})}= -1$} & \multicolumn{1}{c|}{$p_{5}$} & \multicolumn{1}{c|}{$p_{6}$} \\ [2ex]\cline{2-3} 
\end{tabular}
\vspace*{.65 cm}}
\newline
{\setlength{\extrarowheight}{4pt}
\begin{tabular}{ccc}
		$\mathbb{A}_{1}, \mathbb{B}_{2}$ & $a_{\mathbb{B}_{2}}^{(\mathbb{A}_{1}, \mathbb{B}_{2})}= +1$ & $a_{\mathbb{B}_{2}}^{(\mathbb{A}_{1}, \mathbb{B}_{2})}= -1$  \\[2ex]\cline{2-3} 
\multicolumn{1}{c|}{$a_{\mathbb{A}_{1}}^{(\mathbb{A}_{1}, \mathbb{B}_{2})}= +1$} & \multicolumn{1}{c|}{$p_{3}$} & \multicolumn{1}{c|}{$p_{4}$} \\ [2ex]\cline{2-3} 
\multicolumn{1}{c|}{$a_{\mathbb{A}_{1}}^{(\mathbb{A}_{1}, \mathbb{B}_{2})}= -1$} & \multicolumn{1}{c|}{$p_{7}$} & \multicolumn{1}{c|}{$p_{8}$} \\ [2ex]\cline{2-3} 
\end{tabular}}
\vspace*{.65 cm}
\newline
{\setlength{\extrarowheight}{4pt}
\begin{tabular}{ccc}
		$\mathbb{A}_{2}, \mathbb{B}_{1}$ & $a_{\mathbb{B}_{1}}^{(\mathbb{A}_{2}, \mathbb{B}_{1})}= +1$ & $a_{\mathbb{B}_{1}}^{(\mathbb{A}_{2}, \mathbb{B}_{1})}= -1$ \\[2ex]\cline{2-3} 
\multicolumn{1}{c|}{$a_{\mathbb{A}_{2}}^{(\mathbb{A}_{2}, \mathbb{B}_{1})}= +1$} & \multicolumn{1}{c|}{$p_{9}$} & \multicolumn{1}{c|}{$p_{10}$} \\[2ex] \cline{2-3} 
\multicolumn{1}{c|}{$a_{\mathbb{A}_{2}}^{(\mathbb{A}_{2}, \mathbb{B}_{1})}= -1$} & \multicolumn{1}{c|}{$p_{13}$} & \multicolumn{1}{c|}{$p_{14}$} \\[2ex] \cline{2-3} 
\end{tabular}}
\vspace*{.65 cm}
\newline
{\setlength{\extrarowheight}{4pt}
\begin{tabular}{ccc}
			$\mathbb{A}_{2}, \mathbb{B}_{2}$ & $a_{\mathbb{B}_{2}}^{(\mathbb{A}_{2}, \mathbb{B}_{2})}= +1$  & $a_{\mathbb{B}_{2}}^{(\mathbb{A}_{2}, \mathbb{B}_{2})}= -1$  \\[2ex]\cline{2-3} 
\multicolumn{1}{c|}{$a_{\mathbb{A}_{2}}^{(\mathbb{A}_{2}, \mathbb{B}_{2})}= +1$} & \multicolumn{1}{c|}{ $p_{11}$} & \multicolumn{1}{c|}{$p_{12}$} \\[2ex] \cline{2-3} 
\multicolumn{1}{c|}{$a_{\mathbb{A}_{2}}^{(\mathbb{A}_{2}, \mathbb{B}_{2})}= -1$} & \multicolumn{1}{c|}{$p_{15}$} & \multicolumn{1}{c|}{$p_{16}$} \\[2ex] \cline{2-3}
\end{tabular}}
 \newline
\end{table}

\citet{DeBarrosKujala2015} use the term ``consistently connected'' for the general form of non-signaling (or no-disturbance). This condition implies:
\begin{defn}{\label{consistentlyconnected}}
	 A system consisting of random variables $a_{i}^{j}$ is consistently connected if $a_{i}^{j} \sim a_{i}^{j'}$ for every observables $i \in \{1,...,m\}$ that belong to different contexts $j, j'\in \{1, ...,n\}$, this notation means $a_{i}$ has the same distribution in both contexts  $j$ and $j'$. 
\end{defn}

\citet{KujalaDzhafarov2015Necessary} consider CbD as an extended notion of contextuality that allows for inconsistent connectedness (Signaling). In contrast, there are some approaches using the standard notation of random variables that relax the non-signaling condition in the Bell experiment. For example, \citet{BraskChaves2017BellWithCommunication} suggest novel causal interpretations of the CHSH violation allowing communication between two sides of an experiment. Their casual structures can simulate quantum and non-signaling correlations. The comparison of these two approaches warrants further investigation. In particular it is necessary to investigate whether it is possible to associate the directed acyclic graphs representation which is used in this causal framework to the double-indexed random variables and the definition of hidden variables used by CbD. Such an association would provide insights about the meaning of contextuality for inconsistently connected systems in CbD approach.

\subsection{A cyclic contextuality example}\label{Sec:CyclicExamples} 

\citet{KujalaDzhafarov2015Necessary} single out a category of contextual systems with binary random variables and denote them as a cyclic class. 
In this class, each context (or bunch) includes exactly two observables, and each observable is measured in exactly two contexts. The number of observables and the number of contexts are equal to each other and called the rank ($n$) of the system.  The cyclic system of rank 2 forms the simplest contextual scenario in the CbD approach. CbD considers the order effect of projective measurements (in QM) as an example of this rank 2 cyclic system.

It is a common belief that we need at least three measurements to derive the simplest scenario of contextuality in QM \citep{LiangSpekkens2011,KunjwalContextualityThesis2016}. This scenario is designed based on Specker's example of contextuality \citep{Specker2011Thelogic}, which requires three bivalent measurements $\{M_{1},M_{2},M_{3} \}$ that can be measured jointly in pairs but not all at once (i.e. as a triple). In QM, this constraint on the triplewise joint measurement can be attained using three bivalent non-orthogonal measurements (POVMs), for which joint measurability does not imply commutativity \citep{KunjwalGhosh2014MinimalStateDependent}. But here, we focus only on the classical version of Specker's scenario, since CbD does not use the language of quantum observables and cannot therefore discuss scenarios where measurements are non-orthogonal. 

Moreover, we only consider consistently connected systems which means that the associated random variable of a measurement such as $M_{1}$ must have the same distributions in two contexts $\{M_{1}, M_{2}\}$ and $\{M_{1}, M_{3}\}$  (See definition~\ref{consistentlyconnected}). This can be represented as $p_{1}+p_{2} =p_{5}+ p_{6}$ for the probabilities in Table \ref{ThreeBunchesSpecker}.
\begin{table}[!hbtp]
\centering
	\caption{Three bunches in the CbD representation of Specker's scenario. Because of the anti-correlation condition, the probabilities on the diagonal lines are equal to zero and the probabilities on the counter diagonal lines are equal to $0.5$.
	}\vspace*{.30 cm}
	{\setlength{\extrarowheight}{4pt}
\begin{tabular}{ccc}
\scriptsize{Bunch 1} & $a_{M_{2}}^{(M_{1}, M_{2})}=0$ &  $a_{M_{2}}^{(M_{1}, M_{2})}=1$ \\ [2ex]\cline{2-3} 
\multicolumn{1}{c|}{$a_{M_{1}}^{(M_{1}, M_{2})}=0$} & \multicolumn{1}{c|}{$p_{1}=0$} & \multicolumn{1}{c|}{$p_{2}=0.5$} \\ [2ex]\cline{2-3} 
\multicolumn{1}{c|}{$a_{M_{1}}^{(M_{1}, M_{2})}=1$} & \multicolumn{1}{c|}{$p_{3}=0.5$} & \multicolumn{1}{c|}{$p_{4}=0$} \\ [2ex]\cline{2-3} 
\end{tabular} }
\vspace*{.70 cm}
\newline
{\setlength{\extrarowheight}{4pt}
\begin{tabular}{ccc}
\scriptsize{Bunch 2} & $a_{M_{3}}^{(M_{2}, M_{3})}=0$ &  $a_{M_{3}}^{(M_{2}, M_{3})}=1$ \\ [2ex]\cline{2-3} 
\multicolumn{1}{c|}{$a_{M_{2}}^{(M_{2}, M_{3})}=0$} & \multicolumn{1}{c|}{$p_{9}=0$} & \multicolumn{1}{c|}{$p_{10}=0.5$} \\ [2ex]\cline{2-3} 
\multicolumn{1}{c|}{$a_{M_{2}}^{(M_{2}, M_{3})}=1$} & \multicolumn{1}{c|}{$p_{11}=0.5$} & \multicolumn{1}{c|}{$p_{12}=0$} \\ [2ex]\cline{2-3} 
\end{tabular}}
\vspace*{.70 cm}
\newline
{\setlength{\extrarowheight}{4pt}
\begin{tabular}{ccc}
\scriptsize{Bunch 3} & $a_{M_{3}}^{(M_{1}, M_{3})}=0$ &  $a_{M_{3}}^{(M_{1}, M_{3})}=1$ \\ [2ex]\cline{2-3} 
\multicolumn{1}{c|}{$a_{M_{1}}^{(M_{1}, M_{3})}=0$} & \multicolumn{1}{c|}{$p_{5}=0$} & \multicolumn{1}{c|}{$p_{6}=0.5$} \\[2ex] \cline{2-3} 
\multicolumn{1}{c|}{$a_{M_{1}}^{(M_{1}, M_{3})}=1$} & \multicolumn{1}{c|}{$p_{7}=0.5$} & \multicolumn{1}{c|}{$p_{8}=0$} \\[2ex] \cline{2-3} 
\end{tabular}}
 \newline
    \label{ThreeBunchesSpecker}
\end{table}

Specker's scenario aslo requires that the anti-correlation condition be satisfied. \citet{DzhafarovKujala2015ContextualityThreeTypes} present this condition as:
\begin{equation}{\label{CBDNegationSpecker}}
	\begin{split}
Pr[a_{M_{i}}^{(M_{i}, M_{j})}=-a_{M_{j}}^{(M_{i}, M_{j})}]=1,  \quad  i, j \in \{1, 2, 3\}.
\end{split}
\end{equation}

Table \ref{Tab.CBDSpecker} suggests a possible representation of Specker's scenario. Here, we assume $a_{i}^{j'}=a_{i}^{j}$ for any measurement $i$ in two different contexts $j$ and $j'$. Therefore, two random variables $a_{M_{1}}^{(M_{1}, M_{2})}$ and $a_{M_{1}}^{(M_{1}, M_{3})}$ take the same value (e.g. 1). We represent these valuations with horizontal hatching in their corresponding cells. Because of the anti-correlation condition in each pairwise joint measurement, $a_{M_{2}}^{(M_{1}, M_{2})}$ should be $0$.  Furthermore
$a_{M_{2}}^{(M_{2}, M_{3})}$ is $0$ as well, since it belongs to the same measurement $M_{2}$, which are represented by vertical hatching. Continuing this argument, we will reach a contradiction for the value of $a_{M_{3}}^{(M_{2}, M_{3})}$ which is represented by the grid hatching. 
\begin{table}[!hbtp]
\centering
\HatchedCell{start1}{end1}{%
  pattern color=black!23,pattern=horizontal lines}
\HatchedCell{start2}{end2}{%
  pattern color=black!23,pattern=horizontal lines}
\HatchedCell{start3}{end3}{%
  pattern color=black!23,pattern=horizontal lines}
\HatchedCell{start4}{end4}{%
  pattern color=black!23,pattern=vertical lines}  
\HatchedCell{start5}{end5}{%
  pattern color=black!23,pattern=vertical lines}   
\HatchedCell{start6}{end6}{%
  pattern color=black!20,pattern=grid} 

	\caption{Specker's scenario. The horizontal axis represents three measurements and the vertical axis indicates three contexts. The six random variables which can take values $\{0,1\}$ are represented by horizontal and vertical hatchings. The raised contradiction is a proof of contextuality which is represented by the grid hatching.}\label{Tab.CBDSpecker}
	\vspace*{.30 cm}
	{\setlength{\extrarowheight}{5pt}%
\begin{tabular}{cccc}
\cline{1-3}
\multicolumn{1}{!{\hspace*{-0.4pt}\vrule\tikzmark{start1}}c!{\vrule\tikzmark{end1}}}{$a_{M_{1}}^{(M_{1}, M_{2})}$} & \multicolumn{1}{!{\hspace*{-0.4pt}\vrule\tikzmark{start4}}c!{\vrule\tikzmark{end4}}}{$a_{M_{2}}^{(M_{1}, M_{2})}$} & \multicolumn{1}{c|}{} & $\{M_{1}, M_{2}\}$ \\ [1.3ex]\cline{1-3}
\multicolumn{1}{|c|}{} & \multicolumn{1}{!{\hspace*{-0.4pt}\vrule\tikzmark{start5}}c!{\vrule\tikzmark{end5}}}{$a_{M_{2}}^{(M_{2}, M_{3})}$} & \multicolumn{1}{!{\hspace*{-0.4pt}\vrule\tikzmark{start2}}c!{\vrule\tikzmark{end2}}}{$a_{M_{3}}^{(M_{2}, M_{3})}$} & $\{M_{2}, M_{3}\}$ \\[1.3ex] \cline{1-3}
\multicolumn{1}{!{\hspace*{-0.4pt}\vrule\tikzmark{start3}}c!{\vrule\tikzmark{end3}}}{$a_{M_{1}}^{(M_{1}, M_{3})}$} & \multicolumn{1}{c|}{} & \multicolumn{1}{!{\hspace*{-0.4pt}\vrule\tikzmark{start6}}c!{\vrule\tikzmark{end6}}}{$a_{M_{3}}^{(M_{1}, M_{3})}$} & $\{M_{1}, M_{3}\}$ \\[1.3ex] \cline{1-3}
           $M_{1}$            &      $M_{2}$                 &      $M_{3}$                 & 
\end{tabular}}
\end{table}

However, this argument is not matched by the CbD approach since the equality $a_{i}^{j'}=a_{i}^{j}$ violates the double indexing assumption. Instead of this argument, \citet{DzhafarovKujala2015ContextualityThreeTypes} use the concept of coupling to investigate the existence of contextuality in Specker's scenario:
\begin{defn}{\label{cupling}}
	 A coupling of a set of random variables $a_{1},...,a_{n}$ is any jointly set of random variables $b_{1},...,b_{n}$ such that $a_{1} \sim b_{1},...,a_{n} \sim b_{n}$.
\end{defn}
They claim that the system is contextual since there is no maximally connected coupling for the system. In their model, \emph{connection} is defined as a set of random variables (such as: $a_{i}^{1},...,a_{i}^{j}$) with the same observable $i$. And the \emph{maximality} for the coupling of a system of random variables is defined as \citep{DzhafarovKujalaContext-Content2016}:
\begin{defn}{\label{maximallity}}
let $a_{i}^{1},...,a_{i}^{j}$ be a connection of a system of random variables, an associated coupling $b_{i}^{1},...,b_{i}^{j}$ is a maximal coupling if $Pr(b_{i}^{1}=...=b_{i}^{j})$ has the the largest value between all possible couplings. If all the couplings related to the connections of that system are maximal couplings, then the main coupling of the system is maximally connected.
\end{defn}

We will show that this approach has to convert to the above argument (in Table \ref{Tab.CBDSpecker}), since it also requires the equality $a_{i}^{j'}=a_{i}^{j}$. This equality is concluded from the consistently connected (no-disturbance) condition and breaches the double indexing assumption.

\begin{figure}[!hbtp]
	\centering
	\includegraphics[width=.70\textwidth]{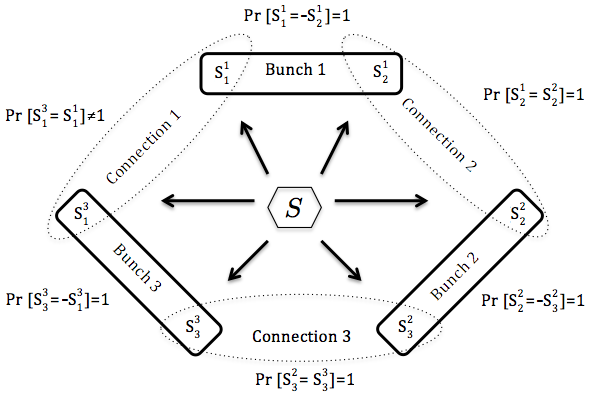}
	\caption{$S$ is a possible maximally connected coupling for the Specker scenario.}
    \label{CouplingSpecker}
\end{figure}

The maximal couplings of three possible connections are constructed in Table \ref{ThreeCouplingsSpecker}. There is a restriction to construct a maximally connected coupling based on the three maximal couplings. As illustrated in Figure \ref{CouplingSpecker}, it is not possible to specify a coupling in which all probabilities are achieved together and still be compatible with the probabilities in the bunches and connections. In this picture, if we associate 0 to the random variable $S_{1}^1$, then $S_{2}^1$ should be 1 because of the anti-correlation condition. Moving clockwise we reach the connection 2, in which two random variables $S_{2}^1$ and $S_{2}^2$ should take a same value and a same distribution because of the consistently connected condition. By moving further clockwise, we will reach a contradiction for the value of $S_{1}^1$ in the connection 1. This restriction on construction of the maximally connected coupling is the proof of the contextuality in the CbD approach.

This proof considers that the two random variables of each connection have the same distribution ($a_{i}^j \sim a_{i}^{j'}$). But since they have the same value (e.g. 1), we can conclude that they are exactly equal to each other $a_{i}^j = a_{i}^{j'}$. This is similar to what we described earlier for the case of non-signaling, where the double indexing notation had to convert to the standard \emph{context-independent} representation of random variables. Here, if we ignore the double indexing scenario, we can remove the three imaginary connections in Figure \ref{CouplingSpecker}, and convert the CbD notation to the standard representation of Specker scenario. This demonstrates that CbD adds extra complexity to the modelling of scenarios like Specker's, without adding new insights to our understanding  of  contextuality.

\begin{table}[!hbtp]
	\caption{Three maximal couplings of the three connections in the CbD representation of Specker's scenario.}
    \label{ThreeCouplingsSpecker}
\centering
\vspace*{.30 cm}
{\setlength{\extrarowheight}{4pt}%
\begin{tabular}{ccc}
\scriptsize{Coupling 1} & $T_{M_{1}}^{(M_{1}, M_{3})}=0$ &  $T_{M_{1}}^{(M_{1}, M_{3})}=1$ \\ [2ex]\cline{2-3} 
\multicolumn{1}{c|}{$T_{M_{1}}^{(M_{1}, M_{2})}=0$} & \multicolumn{1}{c|}{$0.5$} & \multicolumn{1}{c|}{$0$} \\[2ex] \cline{2-3} 
\multicolumn{1}{c|}{$T_{M_{1}}^{(M_{1}, M_{2})}=1$} & \multicolumn{1}{c|}{$0$} & \multicolumn{1}{c|}{$0.5$} \\[2ex] \cline{2-3} 
\end{tabular} }
\vspace*{.70 cm}
\newline
{\setlength{\extrarowheight}{4pt}%
\begin{tabular}{ccc}
\scriptsize{Coupling 2} & $T_{M_{2}}^{(M_{2}, M_{3})}=0$ &  $T_{M_{2}}^{(M_{2}, M_{3})}=1$ \\ [2ex]\cline{2-3} 
\multicolumn{1}{c|}{$T_{M_{2}}^{(M_{1}, M_{2})}=0$} & \multicolumn{1}{c|}{$0.5$} & \multicolumn{1}{c|}{$0$} \\[2ex] \cline{2-3} 
\multicolumn{1}{c|}{$T_{M_{2}}^{(M_{1}, M_{2})}=1$} & \multicolumn{1}{c|}{$0$} & \multicolumn{1}{c|}{$0.5$} \\[2ex] \cline{2-3} 
\end{tabular} }
\vspace*{.70 cm}
\newline
{\setlength{\extrarowheight}{4pt}%
\begin{tabular}{ccc}
\scriptsize{Coupling 3} & $T_{M_{3}}^{(M_{1}, M_{3})}=0$ &  $T_{M_{3}}^{(M_{1}, M_{3})}=1$ \\[2ex] \cline{2-3} 
\multicolumn{1}{c|}{$T_{M_{3}}^{(M_{2}, M_{3})}=0$} & \multicolumn{1}{c|}{$0.5$} & \multicolumn{1}{c|}{$0$} \\[2ex] \cline{2-3} 
\multicolumn{1}{c|}{$T_{M_{3}}^{(M_{2}, M_{3})}=1$} & \multicolumn{1}{c|}{$0$} & \multicolumn{1}{c|}{$0.5$} \\[2ex] \cline{2-3} 
\end{tabular}}
\newline
\end{table}

The other cyclic system of rank 3 in the CbD approach is associated with Leggett-Garg (LG) inequality \citep{LeggettGarg1985flux}. CbD considers a similar structure to Specker's scenario for LG inequality but without the anti-correlation condition \cite{DzhafarovKujalaContext-Content2016}, and interprets the violation of this inequality as contextuality. 

Although the CbD approach can relate bunches to empirical meanings, the coupling itself has been provided with no empirical meaning. \citet[p.~11]{DzhafarovKujalaContext-Content2016} declare that the coupling is merely a mathematical notion: ``If the bunches are assumed to have links to empirical observations, then the couplings can be said to have no empirical meaning. A coupling forms a base set of its own, consisting of itself''. This makes it impossible to \emph{generally} compare the meaning of contextuality in the CbD approach with the other approaches of contextuality in physics. However, in this section we provided mathematical comparisons for cyclic examples, and highlighted some limitations of the CbD approach such as its inability to deal with POVMs and the added extra complexity by its notation. CbD must work to address these limitations before it could be considered a general model of contextuality. 

\section{Conclusion}\label{Sec:Conclution}

\citet{MazurekPusey2016ExperimentaContextuality} suggest an experimental test based on Spekkens operational approach for real situations of inexact operational equivalence. Here, we compared Spekkens' approach with the CbD notation which can also lead to an inexact operational equivalence. This comparison helps us to unify our understandings of contextuality. Especially, it helps us to evaluate the CbD approach and its double indexing notation of random variables. In that regard, we pointed out that there is no clear relation between the double indexing notation and ontic states. We also highlighted that the CbD cannot deal with POVMs. This limits CbD to realize some contextually scenarios in QM, since POVMs allow joint measurability structures which are not compatible with projective measurements alone \citep{KunjwalHeunenFritz2014QuantumRealization}. We illustrated this using a simple example of Specker's scenario, in which CbD cannot represent constraints on the joint measurability that one could realize between POVMs. We mainly explained that the identification of random variables does not add anything to the meaning of contextuality for the systems satisfying non-signaling and no-disturbance conditions (e.g., the Specker scenario), and the double indexing notation \emph{has} to convert to the standard \emph{context-independent} notation for those systems.

In summary, this paper has examined clear differences between the notations used in two extant models of contextuality, drawing attention to both their ontological commitments, and subtle discrepancies in how they approach the parameter independence and non-signalling conditions. The distinctions between the two approaches make it unreasonable to assume that they are equivalent. This shows a need for more research in developing a general understanding of contextuality in QM. We consider this paper a small step towards achieving a more unified understanding of this highly important phenomenon.

\end{document}